\documentclass[prd,twocolumn,eqsecnum,amsfonts,amssymb]{revtex4}

\usepackage{graphicx}

\usepackage{bm}

\newcommand{\beq}{\begin{equation}}
\newcommand{\eeq}{\end{equation}}
\newcommand{\beqs}{\begin{eqnarray}}
\newcommand{\eeqs}{\end{eqnarray}}

\begin{document}

\title{Study of the Question of an Ultraviolet Zero in the Six-Loop 
Beta Function of the O($N$) $\lambda |\vec \phi|^4$ Theory}

\author{Robert Shrock}

\affiliation{C. N. Yang Institute for Theoretical Physics and 
Department of Physics and Astronomy, \\
Stony Brook University, Stony Brook, NY 11794 }

\begin{abstract}

  We study the possibility of an ultraviolet (UV) zero in the six-loop beta
  function of an O($N$) $\lambda |\vec \phi|^4$ field theory in $d=4$ spacetime
  dimensions. For general $N$, in the range of values of $\lambda$ where a
  perturbative calculation is reliable, we find evidence against such a UV zero
  in this six-loop beta function.

\end{abstract}

\maketitle


\section{Introduction}
\label{intro}

A topic of fundamental importance in quantum field theory is the
renormalization-group (RG) behavior of a real $N$-component scalar field
theory in $d=4$ spacetime dimensions.  This theory is defined by the 
path integral
\beq
Z = \int \prod_x [d \vec{\phi}(x)] \, e^{iS} \ , 
\label{zphi}
\eeq
where $S=\int d^4x \, {\cal L}$, and the Lagrangian ${\cal L}$ is given by 
\beq
{\cal L} = \frac{1}{2}(\partial_\mu \vec{\phi}) \cdot
                      (\partial^\mu \vec{\phi})
- \frac{m^2}{2}|\vec{\phi}|^2 - \frac{\lambda}{4!}|\vec{\phi}|^4 \ ,
\label{lag}
\eeq
where $\vec \phi = (\phi_1,...,\phi_N)^T$ is the real scalar field. The
Lagrangian for this $\lambda |\vec{\phi}|^4$ theory is invariant under a global
O($N$) symmetry group whose elements $R$ are rotations acting on
$\vec\phi$. Quantum loop corrections lead to a dependence of the physical
quartic coupling $\lambda = \lambda(\mu)$ on the Euclidean energy/momentum
scale $\mu$ at which this coupling is measured.  The dependence of
$\lambda(\mu)$ on $\mu$ is described by the renormalization-group beta function
of the theory,
\beq
\beta_\lambda = \frac{d\lambda}{dt} \ , 
\label{beta}
\eeq
where $dt=d\ln\mu$ \cite{rg}. At a reference scale $\mu_0$, the quartic
coupling $\lambda(\mu_0)$ is taken to be positive for the stability of the
theory.  The beta function has a series expansion
\beq
\beta_\lambda = \lambda \sum_{\ell=1}^\infty b_\ell \, a^\ell \ , 
\label{betaseries}
\eeq
where 
\beq
a = \frac{\lambda}{(4\pi)^2} \ , 
\label{adef}
\eeq
and $b_\ell$ is the $\ell$-loop coefficient.  The $n$-loop ($n\ell$)
approximation to $\beta_\lambda$ is obtained by replacing $\ell=\infty$ by
$\ell=n$ in the summand in Eq. (\ref{betaseries}), and is denoted as
$\beta_{\lambda,n\ell}$.  Since the one-loop coefficient, $b_1$, is positive,
it follows that $\lambda(\mu) \to 0$ as $\mu \to 0$ in the infrared (IR), i.e.,
the theory is free in this limit. This perturbative result was confirmed by
nonperturbative analyses \cite{nonpert}-\cite{kbook}.

An important question is whether, for the region of $\lambda$ where a
perturbative calculation of the beta function is reliable, the beta function of
this theory exhibits evidence for a zero away from the origin, at some
(positive) value, $\lambda_{UV}$, or equivalently,
$a_{_{UV}}=\lambda_{UV}/(4\pi)^2$.  If so, then this would be an ultraviolet
fixed point (UVFP) of the renormalization group, i.e., as $\mu \to \infty$,
$a(\mu)$ would approach the limiting value $a_{_{UV}}$ (from below).
Correspondingly, if the $n$-loop beta function has one (or more) zero(s) on the
positive real $a$ axis, we denote the one closest to the origin as
$a_{_{UV,n\ell}}$.  A necessary condition for the $n$-loop beta function to
exhibit evidence for a UV zero at a value $a_{_{UV,n\ell}}$, is that the
beta functions calculated to $(n-1)$-loop and $(n+1)$-loop order
should also exhibit respective zeros at values
$a_{_{UV,(n \pm 1)\ell}}$ close to $a_{_{UV,n\ell}}$.  In previous work,
we have investigated this question for general $N$ up to five-loop order in
\cite{lam} and for $N=1$ up to six-loop order in \cite{lam2}, finding evidence
against a UV zero. Our analysis in \cite{lam2} made use of the calculation of
the six-loop beta function for the special case $N=1$ in \cite{kp1}.

In this paper, using the results of the recent calculation of the six-loop beta
function for general $N$ in \cite{kp}, we investigate the question of whether
the beta function for the general O($N$) $\lambda |\vec \phi|^4$ theory
exhibits robust evidence for a UV zero.  We treat the $\lambda |\vec \phi|^4$
theory in isolation and do not try to study possible embeddings in larger
theories.  Since we will investigate the UV properties of the theory, the value
of $m^2$ will not play an important role in our analysis, because $m^2/\mu^2
\to 0$ in the UV limit, independent of the value of $m^2$. For technical
convenience, we take $m^2$ to be positive.

As background, it is worthwhile to inquire whether there is a known quantum
field theory that is IR-free and has a beta function with a UV zero, which is
thus a UVFP of the renormalization group.  The answer to this question is yes;
an example of such a theory is the nonlinear O($N$) $\sigma$ model in
$d=2+\epsilon$ spacetime dimensions, where $\epsilon$ is small. In
Ref. \cite{nlsm}, an exact solution of this theory was calculated in the limit
$N \to \infty$ with $\lambda(\mu) N = \xi(\mu)$ equal to a fixed finite
function of $\mu$.  In this limit, the beta function for this coupling $\xi$
was calculated to be
\beq
\beta_\xi = \frac{d\xi}{dt} = \epsilon \xi 
\Big ( 1 - \frac{\xi}{\xi_{_{UV}}} \Big )
\label{beta_nlsm}
\eeq
for small $\epsilon$, where $\xi_{_{UV}}=2\pi\epsilon$ is a UV fixed point of
the renormalization group.  Hence, in this theory, as the Euclidean
reference scale $\mu$ increases from small values in the IR to large values in
the UV, the running coupling $\xi(\mu)$ increases but approaches the UVFP at
$\xi=\xi_{_{UV}}$ as $\mu \to \infty$.  The question, then, is whether there is
evidence for a similar type of behavior in the O($N$) $\lambda |\vec \phi|^4$
theory in $d=4$ dimensions for a fixed, finite $N$, at the six-loop level.

The organization of the paper is as follows.  In Section
\ref{bjcoefficients_section} we discuss relevant properties of the coefficients
of the beta function.  In Section \ref{zero_section}, after a brief review of
our previous results up to the five-loop level, we present the results of our
new investigation of a possible UV zero in the beta function for general $N$ up
to the six-loop level. Section \ref{pade_section} includes a further analysis
of this question using Pad\'e approximants. Our conclusions are given in
Section \ref{conclusions_section}.  We include some formulas on beta function
coefficients and on discriminants in Appendices \ref{bellcoefficients} and
\ref{discriminants}, and an analysis using Pad\'e approximants of the 
series for an illustrative test function in Appendix \ref{test_function}.


\section{Coefficients of the Beta Function up to Six-Loop Order}
\label{bjcoefficients_section}

\subsection{General} 

It will be convenient to study a beta function that is 
equivalent to $\beta_\lambda$ in (\ref{beta}), namely 
\beq
\beta_a = \frac{da}{dt} = \frac{1}{(4\pi)^2} \, \beta_\lambda \ .
\label{betaadef}
\eeq
This has the series expansion 
\beq
\beta_a = a \sum_{\ell=1}^\infty b_\ell \, a ^\ell \ . 
\label{betaa}
\eeq
The corresponding $n$-loop beta function, denoted $\beta_{a,n\ell}$, is given
by Eq. (\ref{betaa}) with the upper limit of the loop summation index being
$\ell=n$ instead of $\ell=\infty$.  For the tabular listings to be given below,
it is useful to define the scaled coefficients
\beq
\bar b_\ell = \frac{b_\ell}{(4\pi)^\ell} \ . 
\label{bellbar}
\eeq

We also define a reduced beta function with the factor $b_1 a^2$ divided out,
which is thus normalized to unity at $a=0$, namely
\beq
\beta_{a,{\rm red.}} = 1 + \frac{1}{b_1} \sum_{\ell=2}^\infty 
b_\ell \, a^{\ell-1}  \ . 
\label{betar}
\eeq
Analogously with the full beta function, the $n$-loop truncation of this 
reduced beta function is
\beq
\beta_{a,n\ell,{\rm red.}} = 
1 + \frac{1}{b_1} \sum_{\ell=2}^n b_\ell \, a^{\ell-1}  \ . 
\label{betar_nloop}
\eeq
This function serves as a quantitative measure of how much the $n$-loop
beta function differs from the one-loop beta function, since it is equal to the
ratio
\beq
R_{a,n\ell} \equiv \frac{\beta_{a,n\ell}}{\beta_{a,1\ell}} = 
\beta_{a,n\ell,{\rm red.}} \ . 
\label{ran}
\eeq

The one-loop and two-loop coefficients in Eq. (\ref{betaa}) are independent of
the scheme used for regularization and renormalization \cite{bgz74,gross75},
while the $b_\ell$ with $\ell \ge 3$ are scheme-dependent. In the following,
unless otherwise stated, we use the $b_\ell$ coefficients as calculated in
the $\overline{\rm MS}$ scheme \cite{msbar}, since most higher-loop
computations have been performed with this scheme. Effects of scheme 
transformations were discussed in \cite{lam}.  

The one-loop and two-loop coefficients are \cite{bgz74}
\beq
b_1 = \frac{1}{3}(N+8) 
\label{b1}
\eeq
and
\beq
b_2 = -\frac{1}{3}(3N+14) \ . 
\label{b2}
\eeq

In our study of the five-loop beta function of the O($N$) $\lambda |\vec
\phi|^4$ theory in \cite{lam}, we discussed the behavior of the coefficients
$b_\ell$ with $1 \le \ell \le 5$ as functions of $N$, and we refer the reader
to \cite{lam} for this discussion.  Here we briefly review this behavior.  
Where necessary, we generalize $N$ from the positive
integers to the positive real numbers.  Except for $b_1$, which is a polynomial
of degree 1 in $N$, the coefficients $b_\ell$ are polynomials of degree
$\ell-1$ in $N$ \cite{gracey_largeN}, and hence can be written as 
\beq
b_\ell = \sum_{k=0}^{\ell-1} b_{\ell,k}N^k \quad {\rm for} \ \ \ell \ge 2 \ , 
\label{bellform}
\eeq
where the $b_{\ell,k}$ are independent of $N$.
In Table \ref{bbar_nloop_values} we list numerical values of the $b_\ell$ up to
the $\ell=6$ loop level, expressed in terms of the rescaled quantities $\bar
b_\ell$ defined in Eq. (\ref{bellbar}).

The three-loop coefficient, $b_3$, \cite{bgz74,b345}, given in Eq. (\ref{b3})
in Appendix \ref{bellcoefficients}, is positive for all (physical) $N$.  The
four-loop coefficient, $b_4$ \cite{b345,kbook}, is negative for $N=1$ and
decreases (that is, $-b_4$ increases) as $N$ increases up to the value $N =
2143$, at which it reaches a minimum and then increases, passing through zero
to positive values as $N$ increases through the value \cite{nintegral}
\beq
N_{b4z}= 3218.755 \ , 
\label{nb4z}
\eeq
where and below, numerical values are given to the indicated floating-point
accuracy. (In Eq. (\ref{nb4z}) the subscript $b4z$ means ``$b_4$ zero''.) For
larger values of $N$, $b_4$ remains positive.  The five-loop coefficient,
$b_5$, given in Eq. (\ref{b5}) \cite{b345}, is positive for $N=1$ and increases
with increasing $N$, reaching a maximum at $N=374$ and then decreasing, passing
through zero to negative values as $N$ increases through the value
\beq
N_{b5z} = 504.740 \ . 
\label{nb5z}
\eeq
This coefficient remains negative for larger $N$.  

We next discuss the behavior of the six-loop coefficient, $b_6$, recently
calculated in \cite{kp}, as a function of $N$ for $N \ge 1$ (in the
$\overline{\rm MS}$ scheme).  This coefficient is a polynomial of degree 5 in
$N$ involving rational coefficients and Riemann zeta functions $\zeta(s)$ with
$s$ up to 9, where $\zeta(s) = \sum_{n=1}^\infty n^{-s}$.  We refer the reader
to \cite{kp} for the analytic expression, which we have used in our
calculations. Numerically,
\beqs
b_6 &=& (2.10179 \times 10^{-4})N^5 -0.113332N^4-42.4818N^3 \cr\cr
&-&1252.5593N^2 - 10166.274N -23314.7030 \ . \cr\cr
&&
\label{b6num}
\eeqs
At $N=1$, this coefficient $b_6$ is
negative and as $N$ increases, it decreases through negative values 
(i.e., $-b_6$ increases), reaching a minimum and then increasing and 
passing through zero at
\beq
N_{b6z}=800.9505 \ , 
\label{nb6z}
\eeq
and remaining positive for larger $N$. 

With these beta function coefficients now calculated up to six-loop order (with
$b_\ell$ for $3 \le \ell \le 6$ computed in the $\overline{\rm MS}$ scheme), we
can make some comments about them.  The first concerns an
alternating-sign property. The (scheme-independent) coefficients, $b_1$ and
$b_2$, are of opposite sign for all $N$, and the sign of the three-loop
coefficient, $b_3$ is opposite to that of $b_2$ for all $N$.  Over a large
range of $N$ values up to 3218 inclusive, $b_4 < 0$ while for $N$ up to 504
inclusive, $b_5 > 0$.  Additionally, for $N$ up to 800, $b_6 < 0$.  Thus, in
the interval $1 \le N \le 504$, the signs of the $b_\ell$ alternate as a
function of loop order $\ell$ for $1 \le \ell \le 6$.  We will comment further
on this below.

A second salient property is that in each one of these coefficients, 
considered as a polynomial in $N$, the magnitudes of the coefficients of terms
of increasing degree in $N$ decrease as a function of the degree.  This is a
relatively mild effect at low loop level but becomes quite prounounced as the
loop level increases.  Thus, in $b_1=(N+8)/3$, the ratio of the magnitude of 
the term proportional to $N$ to the constant term is $1/8$, while for 
$b_5$, the ratio of the magnitude of the coefficient of the $N^4$ term to that
of the constant term is $(2.57 \times 10^{-3})/(2.004 \times 10^3) = 1.28
\times 10^{-6}$ and for $b_6$, the ratio of the coefficient of the $N^5$ term
to that of the constant term is $(2.10 \times 10^{-4})/(2.33 \times 10^4) =
0.901 \times 10^{-8}$.  

A third property is that in $b_4$, $b_5$, and $b_6$, the coefficient of the
term of highest degree in $N$ is opposite in sign relative to the constant
term.  This property, combined with the second property, means that, as $N$
increases from 1, each of these coefficients passes through zero and reverses
in sign at quite large values of $N$, namely the values $N_{b4z}$, $N_{b5z}$,
and $N_{b6z}$ as given in Eqs.  (\ref{nb4z}), (\ref{nb5z}), and (\ref{nb6z}).
In turn, this means that the asymptotic large-$N$ behavior of these
coefficients only sets in for very large $N$. From general analyses, it has
been concluded that coefficients in perturbative series expansions of
quantities in this $\lambda |\vec\phi|^4$ theory in powers of $a$ at $O(a^n)$
grow asymptotically for large $n$ as a factorial, $\sim n!$ (with additional
factors including $a^n n^b$, where $a$ and $b$ are constants)
\cite{zjbook,asymptotic,kp}. Given the fact that higher-order terms are
scheme-dependent, one understands that this is the generic behavior. This
property underlies the proof that perturbative power series expansions in this
theory are only asymptotic expansions instead of Taylor series expansions with
finite radii of convergence. Here, at least in the commonly used $\overline{\rm
MS}$ scheme, since $b_4$, $b_5$, and $b_6$ vanish for respective large values
of $N$, one must go to much larger values of $N$ before this asymptotic growth
applies.  Fortunately, this is not a complication for our study of a possible
UV zero of the beta function because a very simple analysis applies in the
large-$N$ limit, as will be discussed below.


\section{Zeros of the Beta Function}
\label{zero_section}

\subsection{General}
\label{zero_general_subsection}

In this section we proceed to the main object of this paper, namely the
investigation of a possible UV zero of the six-loop beta function of the 
O($N$) $\lambda |\vec \phi|^4$ theory.  The beta function of this theory
has a double zero at the origin, $a=0$, which is an IR fixed point of the
renormalization group.  In general, the condition that the
$n$-loop beta function, $\beta_{a,n\ell}$, has a zero away from the origin
$a=0$ is the equation of degree $n-1$ in $a$, 
\beq
\sum_{\ell=1}^n b_\ell \, a^{\ell-1} = 0 \ . 
\label{betar_nloop_zero}
\eeq
Here and below, unless otherwise indicated, we use the $b_\ell$ with $\ell \ge
3$ from the calculations up to six-loop order in the $\overline{\rm MS}$ scheme
\cite{kp}. The roots of Eq. (\ref{betar_nloop_zero}) depend on the $n-1$ ratios
$b_\ell/b_1$, $2 \le \ell \le n$.  The investigation of zeros of
$\beta_{a,n\ell}$ away from the origin thus amounts to the study of the zeros
of the reduced $n$-loop beta function, $\beta_{a,n\ell,{\rm red.}}$, defined in
Eq. (\ref{betar_nloop}).  Although only one of the roots of the equation
(\ref{betar_nloop_zero}), or equivalently, $\beta_{a,n\ell,{\rm red.}}=0$, will
be relevant for our analysis, it will be useful to characterize the full set of
roots. A valuable quantity for this purpose is the discriminant of the equation
(\ref{betar_nloop_zero}), denoted $\Delta_{n-1}(b_1,b_2,...,b_n)$
\cite{disc}. We record some relevant definitions and formulas on discriminants
in Appendix \ref{discriminants}.


\subsection{Zeros of the $n$-Loop Beta Function for $2 \le n \le 5$} 
\label{specific_zero_subsection}

Before presenting our new calculations, we briefly summarize some relevant 
results that we have obtained in \cite{lam} concerning possible UV zeros
of the beta function of the general O($N$) $\lambda |\vec \phi|^4$ theory up 
to the five-loop level. 

Because $b_1$ and $b_2$ are of opposite sign, the two-loop beta function, 
$\beta_{a,2\ell}$, has a a UV zero for all physical 
$N$ (i.e., $N \ge 1$). This UV zero occurs at $a=a_{_{UV,2\ell}}$, where 
\beq
a_{_{UV,2\ell}} = -\frac{b_1}{b_2} = \frac{N+8}{3N+14} \ . 
\label{auv_2loop}
\eeq
As $N$ increases from 1 to $\infty$, $a_{_{UV,2\ell}}$ decreases monotonically
from 9/17 to $1/3$.  As noted above, one must examine 
higher-loop results to judge whether this two-loop zero is a robust,
reliable prediction of perturbation theory or whether, on the contrary, it 
occurs at too large a value of $a$ (equivalently, $\lambda$) to be a 
reliable prediction. 

At the three-loop level, the condition that $\beta_{a,3\ell}=0$ at a nonzero
value of $a$ is that $b_1+b_2a +b_3a^2=0$.  This equation does not have any
physical solutions, but instead, two complex-conjugate solutions, for all
physical $N$. This result follows from the fact that the
discriminant (given explicitly as Eq. (3.6) in \cite{lam}) is negative-definite
for all physical values of $N$.

We investigated how robust this conclusion is to scheme transformations in
\cite{lam}.  A natural approach is to devise a scheme transformation as
specified in \cite{sch}-\cite{sch2} that renders $b_3'=0$ in the transformed
scheme.  We showed, however, that although, by construction, the resultant
three-loop beta function in this transformed scheme would be equal to the
two-loop beta function and would hence have a UV zero at $a'_{_{UV,3\ell}} =
a_{_{UV,2\ell}} = -b_1/b_2$, the four-loop and five-loop beta functions in this
transformed scheme do not yield UV zeros close to this value (see Table III in
\cite{lam}).  For example, for $N=1$, while $a'_{_{UV,3\ell}}=
a'_{_{UV,2\ell}}=0.5294$, the zero in the scheme-transformed four-loop beta
function occurs at quite a different value, $a'_{_{UV,4\ell}}=0.1917$, and the
five-loop beta function in this transformed scheme has no physical UV zero.
Similar results hold for other values of $N$. 

At the four-loop level, as the $n=4$ special case of
Eq. (\ref{betar_nloop_zero}), the equation for $\beta_{a,4\ell}=0$ with
$a \ne 0$ is $b_1+b_2a+b_3a^2+b_4a^3=0$.  The properties of the solutions to
this equation are determined by the discriminant
$\Delta_3(b_1,b_2,b_3,b_4)$ given by Eqs. (\ref{deltam3}) and (\ref{cbrel}) in 
Appendix \ref{discriminants}. This is negative for all physical $N$, and hence
these solutions consist of one real value and a complex-conjugate pair of
values of $a$.  In \cite{lam} we showed that for $N$
in the range $1 \le N < N_{b4z}$, the real root is positive, so that the
four-loop beta function has a physical UV zero, $a_{_{UV,4\ell}}$,  
but for $N > N_{b4z}$, this real root becomes negative, so
that this four-loop beta function has no physical UV zero.  Values of
$a_{_{UV,4\ell}}$ for a large range of values of $N$ are listed in Table 
\ref{auv_nloop_values}.

At the five-loop level, the condition for a zero of $\beta_{a,5\ell}$ with $a
\ne 0$ is obtained from Eq. (\ref{betar_nloop_zero}) with $n=5$ and is the
quartic equation $b_1+b_2a+b_3a^2+b_4a^3+b_5a^4=0$.  The discriminant of this
equation, $\Delta_4 \equiv \Delta_4(b_1,b_2,b_3,b_4,b_5)$, is given by
Eqs. (\ref{Deltam_matrix}) and (\ref{sp4pp4}) with (\ref{cbrel}), in Appendix
\ref{discriminants}.  For physical $N$, this discriminant is positive for $1
\le N < N_{\Delta_4 z}$, where $N_{\Delta_4 z} = 493.0957$ \cite{nintegral} and
negative for larger $N$. From this information or the equivalent analysis of
$b_5 \Delta_4$ in \cite{lam}, one then determines the nature of the roots of
the above quartic equation. For values of $N$ from 1 to 493, the five-loop beta
function has no physical UV zero. For larger values of $N$, the quartic
equation has two real positive roots (and a complex-conjugate pair of roots),
and the smaller of these is $a_{_{UV,5\ell}}$. This is listed in Table
\ref{auv_nloop_values}.  For the interval of $N$ in which both the four-loop
and five-loop beta functions have UV zeros, namely $494 \le N \le 3218$, these
zeros, $a_{_{UV,4\ell}}$ and $a_{_{UV,5\ell}}$ are not close to each other.
The values of $a_{_{UV,4\ell}}$ and $a_{_{UV,5\ell}}$ are only approximately
equal if $N$ is close to $N_{b5z}$, so that $b_5=0$ and $\beta_{a,5\ell} =
\beta_{a,4\ell}$, whence $a_{_{UV,4\ell}}$ and $a_{_{UV,5\ell}}$ are
automatically equal. As will be discussed next, in this small region of $N$
close to $N_{b5z}$ where $a_{_{UV,4\ell}} \simeq a_{_{UV,5\ell}}$, these are
not approximately equal to $a_{_{UV,6\ell}}$, as would be expected if this were
a reliably indication of a UV zero in the full beta function. For example, as
indicated in Table \ref{auv_nloop_values}, at $N=500$, where
$a_{_{UV,4\ell}}=0.07341$, close to $a_{_{UV,5\ell}}=0.08045$, these values are
not close to the six-loop value, $a_{_{UV,6\ell}}=0.03074$.


\subsection{Zeros of $\beta_{a,6\ell}$}
\label{betar_6loop_zero_section}

We now present our new results from our investigation of a possible UV zero in
the six-loop beta function of the O($N$) $\lambda |\vec \phi|^4$ theory.  The
condition for a zero of $\beta_{a,6\ell}$ with $a \ne 0$ is the special case of
Eq. (\ref{betar_nloop_zero}) with $n=6$, namely, the quintic equation
$b_1+b_2a+b_3a^2+b_4a^3+b_5a^4+ b_6a^5=0$.  The discriminant, $\Delta_5 \equiv
\Delta_5(b_1,b_2,b_3,b_4,b_5,b_6)$, of this equation is given by
Eqs. (\ref{Deltam_matrix}) and (\ref{sp5pp5}) with (\ref{cbrel}), in the
Appendix \ref{discriminants}.  This discriminant is negative in the interval $1
\le N \le 760.24$, positive for the physical values $761 \le N \le 892$, and
negative for $N > 892.218$ \cite{nintegral}.  We find that the quintic equation
above has a real positive root in the interval $1 \le N \le 892$, but no such
physical root for $N \ge 893$.  Values of the real positive root are listed in
Table \ref{auv_nloop_values}.

A necessary condition for a perturbative calculation of the beta function 
to be reliable is that the fractional change
\beq
\bigg | \frac{\beta_{a,n\ell}-\beta_{a,(n-1)\ell}}{\beta_{a,n\ell}} \bigg | 
\label{betachange}
\eeq
should generally decrease as the loop order $n$ increases, at least away from a
zero of $\beta_{a,n\ell}$.  Another necessary condition for the reliability of a
result on a zero of the $n$-loop beta function, $\beta_{a,(n-1)\ell}$, is that when one
calculates the beta function to the next higher-loop order, viz.,
$\beta_{a,n\ell}$, the zero should still be present and its value should not
shift very much.  For the specific case at hand, where we are investigating a
possible UV zero of the beta function, this condition is that the fractional shift
\beq
\frac{|a_{_{UV,n\ell}}-a_{_{UV,(n-1)\ell}}|}{a_{_{UV,n\ell}}}
\label{achange}
\eeq
should be small.  Our new calculations extend our previous findings, showing to
the six-loop order that these two necessary conditions are not satisfied for
this theory.  In much of this interval $1 \le N \le 892$ where the six-loop
beta function $\beta_{a,6\ell}$ has a UV zero, the five-loop beta function
$\beta_{a,5\ell}$ does not have any UV zero.  In the interval $N \ge 893$,
$\beta_{a,5\ell}$ has a UV zero, but $\beta_{a,6\ell}$ does not, and,
furthermore, the five-loop UV zero, $a_{_{UV,5\ell}}$, is quite different from
the four-loop value, $a_{_{UV,4\ell}}$.  For example, as is evident in Table
\ref{auv_nloop_values}, for $N=2000$, $a_{_{UV,5\ell}}=0.01231$, almost a
factor of ten smaller than the four-loop value, $a_{_{UV,4\ell}}=0.1054$.  In
the small region of $N$ close to $N_{b5z}$ where $a_{_{UV,4\ell}} \simeq
a_{_{UV,5\ell}}$, these are not approximately equal to $a_{_{UV,6\ell}}$, as
would be expected if this were a reliably indication of a UV zero in the full
beta function. For example, as indicated in Table \ref{auv_nloop_values}, at
$N=500$, where $a_{_{UV,4\ell}}=0.07341$, close to $a_{_{UV,5\ell}}=0.08045$,
these values are not close to the six-loop value,
$a_{_{UV,6\ell}}=0.03074$. For the limited interval where both
$\beta_{a,5\ell}$ and $\beta_{a,6\ell}$ have UV zeros, the five-loop and
six-loop values $a_{_{UV,5\ell}}$ and $a_{_{UV,6\ell}}$ are not very close to
each other.  The only exception to this is in the immediate vicinity of $N$
around the special value $N_{b6z}$ where $b_6=0$; at this point,
$\beta_{a,6\ell}=\beta_{a,5\ell}$, so it is automatic that
$a_{_{UV,6\ell}}=a_{_{UV,6\ell}}$.  Finally, for larger $N$, the general
analysis given in \cite{lam} and briefly reviewed below shows the absence of a
UV zero.

\begin{figure}
  \begin{center}
    \includegraphics[height=8cm,width=6cm]{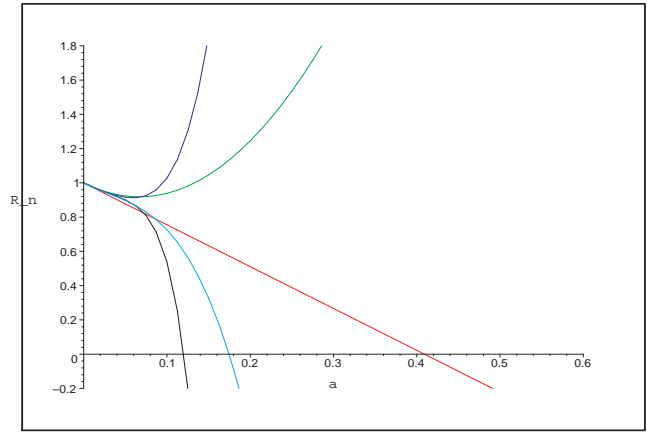}
  \end{center}                             
\caption{\footnotesize{Plot of the ratio $R_n \equiv R_{a,n}$ defined in
    Eq. (\ref{ran}), as a function of $a$, for
    $N=10$ and (i) $n=2$ (red), (ii) $n=3$ (green), (iii) $n=4$ (cyan), (iv)
    $n=5$ (blue), and (v) $n=6$ (black) (colors in online version). Along a
    counterclockwise path around the point $(a,R_n)=(0,1)$ starting at 
    the point $(a,R_n)=(0.1,0)$, the curves are for 
    $n=6$, $n=4$, $n=2$, $n=3$, and $n=5$.}}
\label{lam_betareduced_6loop_Neq10_fig}
\end{figure}

Another way of understanding the absence of a UV zero is by plotting the
reduced $n$-loop beta function, which is equal to the ratio $R_n$ given in
Eq. (\ref{ran}) measuring the relative agreement between the beta functions at
adjacent-loop orders.  In \cite{lam2} in the case $N=1$ we showed these curves
up to the six-loop level, and here we show them for an illustrative higher
value, $N=10$, in Fig.  \ref{lam_betareduced_6loop_Neq10_fig}.  One sees that
the $R_n$ ratios for adjacent values of $n$ ranging from $2 \le n \le 6$ behave
quite differently and do not exhibit the sort of agreement with each other that
one would expect if the beta function had a reliably calculable UV zero.

It is not necessary to carry out specific searches for a UV in the beta
function for large $N$, because in this regime we can apply a more general type
of analysis.  This was done in \cite{lam} and showed the absence of a UV zero
in the $\lambda |\vec \phi |^4$ theory for $N \gg 1$.  As in \cite{lam}, we
define the limit 
\beq
N \to \infty \ , \ \ {\rm with} \ x(\mu) \equiv Na(\mu) \ {\rm a \ finite \
  function \ of} \ \mu. 
\label{ln}
\eeq
This is denoted as the LN limit, with the symbol $\lim_{LN}$.  The two
scheme-independent coefficients, $b_1$ and $b_2$, are both polynomials of
degree 1 in $N$, and the higher-loop coefficients $b_\ell$ are polynomials of
degree $\ell-1$ in $N$ \cite{gracey_largeN}, as indicated in
Eq. (\ref{bellform}). Thus, one can write $b_1 = b_{1,1}N + b_{1,0}$, where
$b_{1,1}=1/3$ and $b_{1,0} = 8/3$.  We extract the leading-$N$ factors and
define
\beq
\hat b_\ell =\lim_{LN} \frac{b_\ell}{N^{\ell-1}} \quad {\rm for} \ \ \ell 
\ge 2 \ . 
\label{bellhat}
\eeq
so that these $\hat b_\ell$ with $\ell \ge 2$ are finite in the large-$N$
limit.  The explicit values of the $\hat b_\ell$ follow from the expressions
for the $b_\ell$ and are
\beq
\hat b_2 = -1 \ , 
\label{b2hat}
\eeq
\beq
\hat b_3 = \frac{11}{72} =0.152778 \ , 
\label{b3hat}
\eeq
\beq
\hat b_4 = \frac{5}{3888} = 1.2860 \times 10^{-3} \ , 
\label{b4hat}
\eeq
\beq
\hat b_5 = \frac{13}{62208} - \frac{\zeta(3)}{432} = -(2.57356 \times 10^{-3})
\ ,
\label{b5hat}
\eeq
and
\beqs
\hat b_6 &=& \frac{29}{933120} + \frac{11}{19440}\zeta(3) -
\frac{\zeta(4)}{2160} \cr\cr
&=& 2.10179 \times 10^{-4} \ . 
\eeqs
(where $\zeta(4)=\pi^4/90$.)

Since the LN limit is defined so that $x(\mu)$ is a finite
function of $\mu$, the appropriate beta function that is finite in this limit
is
\beqs
\beta_x & = & \frac{dx}{dt} = \lim_{LN} N \beta_a \cr\cr
& = & x^2 \bigg [ b_{1,1} + \frac{1}{N} \sum_{\ell=2}^\infty 
\hat b_\ell \, x^{\ell-1} \bigg ] \ . 
\label{betax}
\eeqs
The $n$-loop beta function in the LN limit, denoted $\beta_{x,n\ell}$, is
defined via Eq. (\ref{betax}) with the upper limit on the sum being $\ell=n$
rather than $\ell=\infty$.  From Eq. (\ref{betax}), is it clear that in the LN
limit \cite{lam}, for any given loop order $n$, $\beta_{x,n\ell}$ has no UV
zero $x_{_{UV,n\ell}}$, since
\beq
\lim_{LN} \frac{1}{N} \sum_{\ell=2}^n \hat b_\ell \, x^{\ell-1} = 0 \ .
\label{lnsumzero}
\eeq
Hence, in the $N \to \infty$ limit, as $\mu$ increases, $x(\mu)$ increases,
eventually exceeding the range of values where the perturbative $n$-loop
expansion of $\beta_{x,n\ell}$ is reliable.  This result in the LN limit agrees
with our specific calculations up to the six-loop level for large finite values
of $N$ as shown in Table \ref{auv_nloop_values}.  For example, for $N=10^4$
(chosen to be larger than $N_{b4z}$, $N_{b5z}$, and $N_{b6z}$), the three-loop,
four-loop, and six-loop beta functions have no UV zero, and although the
five-loop beta function $\beta_{a,5\ell}$ has a UV zero, at
$a_{_{UV,5\ell}}=0.003460$, it is a factor of 100 smaller than the two-loop
value, $a_{_{UV,2\ell}}=0.3334$. Thus, neither of the necessary criteria for a
reliably calculable UV zero of the six-loop beta function is satisfied here. 


\section{Analysis With Pad\'e Approximants}
\label{pade_section}


\subsection{General} 

In the search for a possible UV zero of the six-loop beta function of the
O($N$) $\lambda |\vec \phi|^4$ theory, it is also instructive to calculate and
analyze Pad\'e approximants (PAs) to this function. Moreover, these
approximants can be used to investigate the general analytic structure of the
beta function. Since the zero in question would occur away from the origin in
coupling-constant space, it is convenient to extract an overall prefactor of
$b_1 a^2$ and compute Pad\'e approximants to the reduced beta function,
$\beta_{a,n\ell,{\rm red.}}$ defined in Eq. (\ref{betar_nloop}). Our six-loop
results on a possible UV zero for this O($N$) $\lambda |\vec \phi|^4$ theory
extend our previous studies of Pad\'e approximants to the beta function that
were carried out up to the five-loop level for general $N$ in \cite{lam} and up
to the six-loop level for $N=1$ in \cite{lam2}.

For a function $f(a)$ satisfying $f(0)=1$, with a finite series
expansion about $a=0$ given by $f(a) = 1 + \sum_{s=1}^{n-1} c_s a^s$, the $[p,q]$
Pad\'e approximant is the rational function 
\beq
[p,q] = \frac{\sum_{j=0}^p N_j a^j}{\sum_{k=0}^q D_k a^k} \ ,
\label{pqpade}
\eeq
with polynomials in the numerator and denominator of degree $p$ and $q$,
respectively, where $p+q=n-1$ and $N_0=1=D_0$ \cite{pade}.  The 
coefficients $N_j$ with $j=1,...,p$ and $D_k$ with $k=1,...,q$ are 
determined by the $m$ coefficients $c_1,...,c_{n-1}$, so that the Taylor series
expansion of the $[p,q]$ Pad\'e approximant about $a=0$ matches the
corresponding expansion of $f(a)$ up to its maximal order, $O(a^{n-1})$. 
For our application, $f(a) = \beta_{a,n\ell,{\rm red.}}$ and
$c_s=b_{s+1}/b_1$ for $1 \le s \le n-1$. 

We recall some general properties of these Pad\'e approximants.  The $[n-1,0]$
PA to $\beta_{a,n\ell,{\rm red.}}$ is this function itself, i.e.,
\beq
[n-1,0] = \beta_{a,n\ell,{\rm red.}} \ . 
\label{polynomialpade}
\eeq
Since we have already analyzed the zeros of $\beta_{a,n\ell,{\rm red.}}$ above,
we do not discuss the $[n-1,0]$ approximants here. Moreover, the $[0,n-1]$ PA
approximant has no zeros and hence is not useful for investigating a possible
UV zero in the beta function.  Thus, for the purpose of investigating a
possible UV zero in the beta function, we shall use the $[p,q]$ PAs with $p \ne
0$ in addition to the analysis that we have already carried out for
$\beta_{a,n\ell,{\rm red.}}$.  

In order for a zero of Pad\'e approximant to $\beta_{a,n\ell,{\rm red.}}$ to be
physically meaningful, (i) it must occur on the positive real $a$ axis, and
(ii) calculations of Pad\'e approximants to the (reduced) $n$-loop beta
functions with different loop orders should yield approximately the same value
for this zero. Furthermore, (iii) if the Pad\'e approximant has a pole on the
positive real $a$ axis, this pole must not occur closer to the origin than the
zero. This is clear, since if there were such a pole, then as $\mu$ increases
from small values in the IR to large values in the UV and $a(\mu)$ increases
from the vicinity of the origin, it would approach the pole before it reached
the zero.  In order for a zero of a $[p,q]$ Pad\'e approximant to be considered
physically meaningful, one might also consider imposing a stricter condition,
namely that this zero must occur within the disk in the complex $a$ plane in
which the PA has a convergent Taylor series expansion.  Since the radius of
this disk is determined by the real pole or pair of complex-conjugate poles
closest to the origin, this condition would be that, in addition to properties
(i)-(iii), the zero must be closer to the origin than any pole(s), even if a
pole occurs on the negative real axis or if the PA has complex-conjugate pairs
of poles.  However, we will not have to consider imposing this last condition,
since the zeros of PAs that we find do not satisfy the first three conditions.
It should also be noted that $[p,q]$ Pad\'e approximants in which both $p$ and
$q$ are nonzero may exhibit nearly or exactly coincident pairs of zeros and
poles.  This type of behavior typically occurs if one tries to approximate
(from the series expansion) a function that has fewer than $p$ zeros and $q$
poles with a $[p,q]$ Pad\'e approximant.  As will be evident from our results
below, a number of the higher-order Pad\'e approximants that we compute exhibit
poles and zeros at points that are quite close to each other. In this case, it
is expected that one may ignore these zero-pole pairs; i.e., the approximant is
indicating that the actual function does not have a zero or pole at the nearly
coincident points.

As was noted above, the one-loop and two-loop coefficients in the beta function
have the opposite sign, and for a large range of values of $N$, this sign
alternation also holds for the higher-loop coefficients up to the highest-loop
level to which they have been calculated, namely the six-loop level (in the
$\overline{\rm MS}$ scheme).  A function with a pole on the negative real axis
could produce this type of sign alternation in a series expansion.  For this
reason, we will analyze the $[p,q]$ Pad\'e approximants to $\beta_{a,n\ell,{\rm
    red.}}$ with $q \ne 0$ to investigate indications of a possible pole in
this function on the negative $a$ axis.

The coefficients $N_j$ and $D_k$ in the $[p,q]$ Pad\'e approximant
(\ref{pqpade}) are, themselves, rational functions of the coefficients $b_\ell$
with $1 \le \ell \le n$. For example, the three-loop reduced beta function is
given by the $n=3$ special case of Eq. (\ref{betar_nloop}), namely
$\beta_{a,3\ell,{\rm red.}} = 1 + (b_2/b_1)a + (b_3/b_1)a^2$, which is
identical to the [2,0] PA.  This function $\beta_{a,3\ell,{\rm
    red.}}$ has no physical zeros, but instead a complex-conjugate pair of
zeros for all $N \ge 1$ \cite{lam}. The [1,1] PA to this function is
\beq
[1,1] = \frac{1 + \Big ( \frac{b_2^2-b_1b_3}{b_1b_2} \Big ) a}
{1 - \Big ( \frac{b_3}{b_2}\Big ) a } \ .
\label{pade11_bform}
\eeq
This [1,1] PA has no physical zeros \cite{lam}; the formal zero is given by
\begin{widetext}
\beq a_{[1,1]_{\rm zero}} = \frac{b_1b_2}{b_1b_3-b_2^2} =
-\frac{72(N+8)(3N+14)}{33N^3+(538+480\zeta(3))N^2+
(4288+5952\zeta(3))N+(9568+16896\zeta(3))} \ .
\label{pade11zero}
\eeq
This is manifestly negative for all physical $N$.  This [1,1] PA also has a
pole at 
\beq
a_{[1,1]_{\rm pole}} = \frac{b_2}{b_3}
= -\frac{72(3N+14)}{33N^2+(922+480\zeta(3))N+(2960+2112\zeta(3))} \ , 
\label{pade11pole}
\eeq
which is also clearly negative for all physical $N$.  In passing, we note that
the pole occurs closer to the origin than the zero, as is evident from the fact
that the difference is a positive quantity:
\beqs
&& a_{[1,1]_{\rm pole}} - a_{[1,1]_{\rm zero}} = \frac{b_2^3}{b_3(b_2^2-b_1b_3)}
\cr\cr
&=& \frac{5184(3N+14)^3}
{\Big [ 33N^2+(922+480\zeta(3))N+(2960+2112\zeta(3)) \Big ]
\Big [33N^3+(538+480\zeta(3))N^2+(4288+5952\zeta(3))N+
(9568+16896\zeta(3)) \Big ]} \ . \cr\cr
&&
\label{pade11_pole_zero_dif}
\eeqs
\end{widetext}

The [0,2] Pad\'e approximant to $\beta_{a,3\ell,{\rm red.}}$ is
\beq
[0,2] = \frac{1}{1 - \Big ( \frac{b_2}{b_1} \Big ) \, a + 
\frac{(b_2^2-b_1b_3)}{b_1^2} \, a^2 } \ .
\label{pade02_bform}
\eeq
This approximant has poles at
\beq
a_{[0,2]_{\rm pole}} = \frac{b_1[b_2 \pm \sqrt{4b_1b_3-3b_2^2} \ ]}
{2(b_2^2-b_1b_3)} \ .
\label{pade02poles}
\eeq
Similar but progressively more complicated analytic expressions can be given 
for the higher-order $[p,q]$ Pad\'e approximants in terms of the coefficients
$b_n$ and explicitly as rational functions of $N$, but these are sufficient to
illustrate the results. 


\subsection{Analysis for Theory with $N=1$} 

We begin with the case $N=1$.  The six-loop beta function for this case was
calculated in \cite{kp1} and analyzed for a possible UV zero in \cite{lam2}. 
Numerically, 
\beqs
\beta_{a,6\ell} & = & a^2\Big ( 3 - \frac{17}{3}a + 32.5497a^2 -271.6058a^3
\cr\cr
& + & 2848.568 a^4 -34776.131 a^5 \Big ) \ . 
\label{beta_6loop}
\eeqs
The reduced six-loop beta function is thus
\beqs \beta_{a,6\ell,{\rm red.}} & = & 1 - 1.8888889a + 10.8499a^2 -90.53527a^3
\cr\cr 
&+&949.5227a^4 -11592.044a^5 \ .
\label{betar_6loop}
\eeqs
The $[p,q]$ Pad\'e approximants (with $p \ne 0$) to the $n$-loop beta functions
with $3 \le n \le 6$ were calculated and studied in \cite{lam2}. We recall
these here.  Since we are also investigating a possible pole on the negative
real axis here, we calculate and analyze the $[0,q]$ Pad\'e
approximants. At the three-loop level, the $[p,q]$  Pad\'e approximants for
$\beta_{a,3\ell,{\rm red.}}$ (with $q \ne 0$) are
\beq
[1,1] = \frac{1+3.85517a}{1+5.74406a} 
\label{bpade11}
\eeq
and 
\beq
[0,2] = \frac{1}{1+1.88889a - 7.28199a^2} \ . 
\label{pade02_Neq1}
\eeq
At the four-loop level, the PAs for $\beta_{a,4\ell,{\rm red.}}$ are 
\beq
[2,1] = \frac{1+6.45546a-4.91165a^2}{1+8.344345a} \ , 
\label{bpade21}
\eeq
\bigskip
\beq
[1,2] = \frac{1+7.72950a}{1+9.61839a+7.31817a^2} \ , 
\label{bpade12}
\eeq
and
\beq
[0,3] = \frac{1}{1+1.88889a - 7.28199a^2+56.2861a^3} \ . 
\label{pade03_Neq1}
\eeq
Proceeding to the five-loop level, the Pad\'e approximants to 
$\beta_{a,5\ell,{\rm red.}}$ are 
\beq
[3,1] = \frac{1+8.5989a-8.9605a^2+23.2571a^3}{1+10.4879a} \ ,
\label{bpade31}
\eeq
\beq
[2,2] = \frac{1+13.3341a+21.6066a^2}{1+15.2230a+39.51125a^2} \ ,
\label{bpade22}
\eeq
\beq
[1,3] = \frac{1+10.5387a}{1+12.4276a+12.6245a^2-20.4568a^3} \ , 
\label{bpade13}
\eeq
and
\beqs
[0,4] &=& \frac{1}{1+1.88889a - 7.28199a^2+56.2861a^3 - 593.1846a^4} \ . 
\cr\cr
&& 
\label{pade04_Neq1}
\eeqs
Finally, at the six-loop level, the Pad\'e approximants to 
$\beta_{a,6\ell,{\rm red.}}$ are 
\beq
[4,1] = \frac{1+10.3193a-12.2102a^2+41.9233a^3-155.757a^4}
{1+12.2083a} \ ,
\label{bpade41}
\eeq
\beq
[3,2] = \frac{1+17.0166a+45.3789a^2-18.0872a^3}{1+18.9055a+70.2394a^2} \ ,
\label{bpade32}
\eeq
\beq
[2,3] = \frac{1+17.8537a+56.5411a^2}{1+19.7426a+82.9828a^2 +33.0754a^3} \ ,
\label{bpade23}
\eeq
\beqs
[1,4] & = & \frac{1+12.48863a}{1+14.3775a+16.3076a^2-34.6560a^3+ 109.7524 a^4}
\ \cr\cr
& &
\label{bpade14}
\eeqs
and
\begin{widetext}
\beq
[0,5] =  \frac{1}{1+1.88889a - 7.28199a^2+56.2861a^3 - 593.1846a^4 
+ 7408.0652a^5} \ . 
\label{pade05_Neq1}
\eeq
\end{widetext}

We list the real zeros and poles in these Pad\'e approximants in Table
\ref{pade_zp_Neq1}.  In order for these various $[p,q]$ Pad\'e approximants to
the reduced $n$-loop beta functions $\beta_{a,n\ell,{\rm red.}}$ to give
evidence for a UV zero in the actual beta function, or equivalently, in
$\beta_{a,{\rm red.}} = \beta_{a,\infty \ell,{\rm red.}}$, the ones with $p \ne
0$ would have to consistently feature a zero at approximately the same value of
$a$.  Clearly, they do not do this. Many of the approximants have no physical
(real, positive) zero and for the the ones that do, the respective values are
significantly different from each other.  Furthermore, for the loop orders
$n=2, 4, 6$ where the respective $n$-loop beta functions do exhibit UV zeros
(namely $a_{_{UV,2\ell}}=0.5294$, $a_{_{UV,4\ell}}=0.2333$, and
$a_{_{UV,6\ell}}=0.1604$), the corresponding sets of Pad\'e approximants do not
reproduce these zeros.  This is automatic at the two-loop level, since the only
PA other than $\beta_{a,2\ell,{\rm red.}}$ itself is [0,1], which has no
zero. At the $n=4$ loop level, the [1,2] PA has no physical UV zero, and
although the [2,1] has one physical zero, it occurs at $a=1.4543$, six times
larger than the UV zero $a_{_{UV,4\ell}}$ in the four-loop beta
function. Moreover, the perturbative expansion of the [2,1] PA only converges
in the disk $|a| < 0.1198$ whose radius is set by the position of its pole at
$a=-0.1198$, and both the physical zero and the unphysical zero of this [2,1]
PA lie outside this disk.  Similarly, the unphysical zero of the [1,2] PA lies
outside the radius of convergence of the Taylor series expansion of this PA,
which is set by its unphysical pole at $a=-0.1138$.  At the six-loop level,
none of the Pad\'e approximants exhibits a zero near to the UV zero in the
six-loop beta function, at $a_{_{UV,6\ell}}=0.1604$.

Aside from the two-loop level, where the pole in the [0,1] PA always occurs at
minus the value of the zero in the [1,0] PA, in each case where $p \ne 0$ so
that a $[p,q]$ PA has one or more zeros, this approximant has a pole closer to
the origin than the zero(s). Moreover, one can also observe many examples of
nearly coincident zero-pole pairs.  For example, at the six-loop level, the
[4,1] PA has a zero at $a=-0.085$ and a pole at $a=-0.082$, the [3,2] PA has a
zero at $a=-0.074$ and a pole at $a=-0.072$, and the [2,3] PA has a zero at
$a=-0.073$ and a pole at $a=-0.072$, and so forth for other approximants (see
Table \ref{pade_zp_Neq1} for values listed with more digits). 

We may also use these Pad\'e approximants to investigate the possible presence
of a pole in the $n$-loop beta functions.  As noted above, for a large range of
values of $N$, the coefficients $b_n$ alternate in sign.  In general, if the
Taylor series expansion of a function $f(a)$ around $a=0$ has this property of
alternating signs, it can indicate the influence of a pole on the negative $a$
axis. The $[p,q]$ Pad\'e approximants with $q \ne 0$ thus provide a test for a
possible pole in the beta function.  In general, one would expect that if a
pole were present in the full beta function, then for many values of $p$ and $q
\ne 0$, the $[p,q]$ Pad\'e approximant would feature a pole at approximately
the position of the pole in this full beta function.  However, these 
Pad\'e approximants do not do this. Our results in Table \ref{pade_zp_Neq1} do
not yield persuasive evidence for such a pole, although they do not exclude
this possibility. In particular, although the [2,2] PA to
$\beta_{a,5\ell,{\rm red.}}$ and the [1,4] PA to $\beta_{a,6\ell,{\rm red.}}$
both exhibit a pole at $a=-0.301$, this pole is not present in the other
$[p,q]$ Pad\'es to the $\beta_{a,n\ell,{\rm red.}}$ functions with $2 \le n \le
6$.  Furthermore, as was true with a number of the zeros in various Pad\'e
approximants, we find that many poles are members of nearly coincident
zero-pole pairs, indicating that they are not likely to actually be present in
the beta function. 

Summarizing, our Pad\'e analysis of the $N=1$ $\lambda \phi^4$ theory does not
give robust evidence for a UV zero of the beta function.  We have used it also
to probe for a possible pole at negative $a$ in the beta function, and have not
found compelling evidence of this either.


\subsection{Cases with Larger $N$} 

It is also valuable to carry out a corresponding calculation and analysis of
Pad\'e approximants for the (reduced) six-loop beta function of the $\lambda
|\vec \phi|^4$ theory with higher values of $N$.  We have performed this
study.  We show the resultant zeros and poles of Pad\'e approximants for the
illustrative value $N=10$ in Table \ref{pade_zp_Neq10}.  These are
qualitatively similar to our results for the theory with $N=1$, and lead to the
same conclusions.  We find similar results for other values of $N$. 

Thus, from our calculation and analysis of Pad\'e approximants to the $n$-loop
beta function up to the $n=6$ loop level, we add to the evidence that we
obtained from the analysis of the zeros of $\beta_{a,n\ell,{\rm red.}}$ against
a reliably calculable UV zero in the beta function of the $\lambda
|\vec{\phi}|^4$ theory.



\subsection{Extensions} 
\label{extensions_section} 

Here we have considered a the O($N$) $\lambda |\vec \phi|^4$ scalar field
theory in isolation. This type of analysis complements studies of more
complicated theories with scalar, fermion, and gauge fields and hence multiple
(quartic, Yukawa, and gauge) couplings.  The beta functions in the latter
theories involve not only powers of single couplings, but also terms containing
products of different couplings, and, understandably, have not been calculated
in general to an order as high as six loops. The renormalization-group behavior
of theories with scalar and fermion fields have been studied both
perturbatively \cite{cel} and nonperturbatively \cite{yuk}. For fully
nonperturbative analyses, the lattice formulation has provided a powerful
tool. Recent studies using perturbatively calculated beta functions that have
found RG fixed points include \cite{cp3}, motivating continued interest in the
phenomenon of asymptotic safety in these multiple-coupling theories.


\section{Conclusions} 
\label{conclusions_section}

In this work we have investigated whether the beta function for the O($N$)
$\lambda |\vec \phi |^4$ theory in $d=4$ spacetime dimensions exhibits evidence
for an ultraviolet zero, using the six-loop beta function recently calculated
in \cite{kp}.  For the range of quartic coupling $\lambda$, or equivalently,
$a=\lambda/(16\pi^2)$, where a perturbative calculation is reliable, we do not
find evidence for such a UV zero.  This conclusion is in accord with, and
extends, our five-loop analysis in \cite{lam} for general $N$ and our six-loop
analysis in \cite{lam2} for $N=1$. Our methods include both analysis of the
zeros of the six-loop beta function itself and calculation and study of the
zeros of Pad\'e approximants.  Our conclusion provides further support for the
modern view of the O($N$) $\lambda |\vec{\phi}|^4$ theory as an effective field
theory that is applicable only over a restricted range of momentum scales
$\mu$.  In view of the alternating nature of the series expansion for the beta
function, we have also used Pad\'e approximants to investigate possible
indications for a pole in the beta function at negative $a$, but we have not
found persuasive evidence for this.


\begin{acknowledgments}

This research was partially supported by the U.S. NSF Grant 
NSF-PHY-16-1620628. 

\end{acknowledgments}


\begin{appendix}


\section{Beta Function Coefficients at Loop Orders $n=3, \ 4, \ 5$}
\label{bellcoefficients}

In this appendix, for reference, we list the $n$-loop coefficients $b_n$ in the
beta function (\ref{betaa}) for $3 \le n \le 5$, as calculated in the widely
used $\overline{\rm MS}$ scheme. Numerical values of the equivalent
coefficients $\bar b_n = b_n/(4\pi)^n$ are given in Table
\ref{bbar_nloop_values} for a relevant set of values of $N$. 

The coefficient $b_3$ is \cite{bgz74,b345,kbook} 
\beqs
b_3&=&\frac{11}{72}N^2+\bigg ( \frac{461}{108}+\frac{20\zeta(3)}{9}\bigg ) N 
+ \frac{370}{27} + \frac{88\zeta(3)}{9} \ . \cr\cr
&& 
\label{b3}
\eeqs
Numerically, 
\beq
b_3 = 0.15278 N^2 + 6.93976 N + 24.4571 \ , 
\label{b3num}
\eeq
to the indicated floating-point accuracy.  Clearly, for all physical $N$, 
$b_3$ is positive and is a monotonically increasing function of $N$. 

The four-loop coefficient is \cite{b345}
\begin{widetext}
\beqs
b_4 &=& \frac{5}{3888}N^3+\bigg ( -\frac{395}{243} - \frac{14\zeta(3)}{9}
+ \frac{10\zeta(4)}{27} - \frac{80\zeta(5)}{81} \bigg ) N^2 
+\bigg (-\frac{10057}{486} - \frac{1528\zeta(3)}{81} + \frac{124\zeta(4)}{27}
- \frac{2200\zeta(5)}{81} \bigg ) N \cr\cr
&-& \frac{24581}{486} -
\frac{4664\zeta(3)}{81} + \frac{352\zeta(4)}{27} - \frac{2480\zeta(5)}{27} \ .
\label{b4}
\eeqs
\end{widetext}
Numerically,
\beqs
b_4 & = & (1.2860 \times 10^{-3})N^3 - 4.11865N^2 - 66.5621N \cr\cr
    & - & 200.92637 \ .
\label{b4num}
\eeqs

The coefficient $b_5$ is \cite{b345}
\begin{widetext}
\beqs
b_5 & = & \bigg ( \frac{13}{62208} - \frac{\zeta(3)}{432} \bigg ) N^4
+ \bigg ( \frac{6289}{31104} + \frac{26\zeta(3)}{81} - \frac{2\zeta(3)^2}{27}
- \frac{7\zeta(4)}{24} + \frac{305\zeta(5)}{243} - \frac{25\zeta(6)}{81}
\bigg ) N^3 \cr\cr
& + &
\bigg ( \frac{50531}{3888} + \frac{8455\zeta(3)}{486} - \frac{59\zeta(3)^2}{81}
- \frac{347\zeta(4)}{54} + \frac{7466\zeta(5)}{243} - \frac{1775\zeta(6)}{243}
+ \frac{686\zeta(7)}{27} \bigg )N^2 \cr\cr
& + & \bigg ( \frac{103849}{972} + \frac{69035\zeta(3)}{486} +
\frac{446\zeta(3)^2}{81} - \frac{2383\zeta(4)}{54} + \frac{66986\zeta(5)}{243}
- \frac{7825\zeta(6)}{81} + 343\zeta(7) \bigg )N \cr\cr
& + & \frac{17158}{81} + \frac{27382\zeta(3)}{81} + \frac{1088\zeta(3)^2}{27}
- \frac{880\zeta(4)}{9} + \frac{55028\zeta(5)}{81} - \frac{6200\zeta(6)}{27}
+ \frac{25774\zeta(7)}{27} \ .
\label{b5}
\eeqs
Numerically,
\beq
b_5 = -(2.57356 \times 10^{-3})N^4 + 1.152827 N^3 + 72.23315N^2 + 771.20866N
+ 2003.97619 \ .
\label{b5num}
\eeq

\end{widetext} 


\section{Discriminants} 
\label{discriminants}

The analysis of the zeros of $\beta_{a,n\ell}$ requires an analysis of the 
zeros of the equation (\ref{betar_nloop_zero}), of degree $n-1$ 
in the variable $a$, given by Eq. (\ref{adef}). For this purpose, we use the
discriminant.  Given a polynomial equation of degree $m$ in the
variable $a$, $P_m(a)=0$, where $P_m(a) = \sum_{s=0}^m c_s a^s$, we will 
label the $m$ roots as $P_m(a)=0$ as $\{a_1,...,a_m \}$.  The discriminant 
of this equation is \cite{disc}
\beq
\Delta_m \equiv \Big [ c_m^{m-1} \prod_{i < j} (a_i-a_j) \Big ]^2 \ .
\label{deltam}
\eeq
Since $\Delta_m$ is a symmetric polynomial in the roots of the equation
$P_m(a)=0$, the symmetric function theorem shows that it can be written 
as a polynomial in the coefficients of $P_m(a)$ \cite{symfun}, as indicated in
the notation $\Delta_m(c_0,...,c_m)$.  For our purposes in 
analyzing the roots of Eq. (\ref{betar_nloop_zero}), we have
\beq
c_s = b_{s+1} 
\label{cbrel}
\eeq
for $s=0,...,m$.  Since 
Eq. (\ref{betar_nloop_zero}) for the zeros of the $n$-loop beta 
function away from the origin is of degree $m=n-1$ in $a$, its discriminant is 
$\Delta_{n-1}(b_1,b_2,...,b_n)$. 

The discriminant $\Delta_m$ can be calculated in terms of the 
$(2m-1) \times (2m-1)$ Sylvester matrix of $P_m(a)$ and $P_m(a)' = dP(a)/da$, 
proportional to the matrix $S_{P_m,P_m'}$ \cite{disc}: 
\beq
\Delta_m = (-1)^{m(m-1)/2}c_m^{-1}{\rm det}(S_{P_m,P_m'}) \ .
\label{Deltam_matrix}
\eeq
The $m=2$ discriminant is well-known; 
$\Delta_2(c_0,c_1,c_2)=c_1^2-4c_0c_2$. 

For $\Delta_3$, the $S_{P_3,P_3'}$ matrix is
\beq
S_{P_3,P_3'} = \left( \begin{array}{ccccc}
c_3  & c_2  & c_1  & c_0  & 0   \\
 0   & c_3  & c_2  & c_1  & c_0 \\
3c_3 & 2c_2 & c_1  &  0   &  0  \\
  0  & 3c_3 & 2c_2 & c_1  &  0  \\
  0  &   0  & 3c_3 & 2c_2 & c_1 \end{array} \right ) \ , 
\label{sp3pp3}
\eeq
yielding the discriminant 
\beqs
\Delta_3(c_0,c_1,c_2,c_3) & = & c_1^2 c_2^2 - 27c_0^2 c_3^2
- 4(c_0 c_2^3 + c_3 c_1^3) \cr\cr
& + & 18 c_0 c_1 c_2 c_3 \ .
\label{deltam3}
\eeqs

For $\Delta_4$ and $\Delta_5$, the relevant $S_{P_m,P_m'}$ matrices are
\beq
S_{P_4,P_4'} = \left( \begin{array}{ccccccc}
c_4  & c_3  & c_2  & c_1  & c_0  &  0   &  0  \\
 0   & c_4  & c_3  & c_2  & c_1  & c_0  &  0  \\
 0   &  0   & c_4  & c_3  & c_2  & c_1  & c_0 \\
4c_4 & 3c_3 & 2c_2 & c_1  &  0   &  0   &  0  \\
 0   & 4c_4 & 3c_3 & 2c_2 & c_1  &  0   &  0  \\
 0   &  0   & 4c_4 & 3c_3 & 2c_2 & c_1  &  0  \\
 0   &  0   &  0   & 4c_4 & 3c_3 & 2c_2 & c_1 \end{array} \right )
\label{sp4pp4}
\eeq
and
\beq
S_{P_5,P_5'} = \left( \begin{array}{ccccccccc}
c_5  & c_4  & c_3  & c_2  & c_1  & c_0 &  0   &  0    & 0  \\
 0   & c_5  & c_4  & c_3  & c_2  & c_1  & c_0 &  0    & 0  \\
 0   &  0   & c_5  & c_4  & c_3  & c_2  & c_1 & c_0   & 0  \\
5c_5 & 4c_4 & 3c_3 & 2c_2 & c_1  &  0   &  0   &  0   & 0  \\
 0   & 5c_5 & 4c_4 & 3c_3 & 2c_2 & c_1  &  0   &  0   & 0  \\
 0   &  0   & 5c_5 & 4c_4 & 3c_3 & 2c_2 & c_1  &  0   & 0  \\
 0   &  0   &  0   & 5c_5 & 4c_4 & 3c_3 & 2c_2 & c_1  & 0  \\
 0   &  0   &  0   &  0   & 5c_5 & 4c_4 & 3c_3 & 2c_2 & c_1
 \end{array} \right )
\label{sp5pp5}
\eeq

From these matrices we calculate the corresponding discriminants according to
Eq. (\ref{Deltam_matrix}) with (\ref{cbrel}).  At the $n$-loop level, the
relevant equation for a UV zero is Eq. (\ref{betar_nloop_zero}), of degree
$n-1$ in $a$.  It follows from
Eqs. (\ref{cbrel}) and (\ref{Deltam_matrix}) that the disciminant for this
Eq. (\ref{betar_nloop_zero}), namely $\Delta_{n-1}(b_1,...,b_n)$, is a
homogeneous polynomial of degree $2(n-2)$ in the beta function coefficients
$b_\ell$, $1 \le \ell \le n$, i.e., 
\beq
{\rm deg}_{\{b_\ell \}} [\Delta_{n-1}(b_1,...,b_n)] = 2(n-2) \ .  
\label{Delta_degb}
\eeq
This is illustrated at the $n=3$ loop level, by 
$\Delta_2(b_1,b_2,b_3)=b_1^2-4b_1b_3$, at the $n=4$ loop level by 
\beqs
\Delta_3(b_1,b_2,b_3,b_4) &=& b_2^2b_3^2-27b_1^2b_4^2-4(b_1b_3^3+b_4b_2^3)
\cr\cr
&+&18b_1b_2b_3b_4 
\label{Delta3}
\eeqs
and so forth for higher loop order $n$. 


\section{Illustrative Function and Analysis}
\label{test_function}

As an illustration of the effectiveness of Pad\'e approximants in testing for
indications of zeros and poles in a function based on information from its
Taylor series expansion, in this appendix we construct and analyze a test
function using these approximants.  Thus, let us consider the rational function
\beq
f(a) = \frac{1+ra}{1+sa} \ , 
\label{ftest}
\eeq
where $r$ and $s$ are real constants with $s > 0$ and $r \ge 0$.  
This function has a zero at $a=-1/r$ and a pole at $a=-1/s$. It has 
the Taylor series expansion about $a=0$
\beq
f(a) = 1 +(r-s)\sum_{k=1}^\infty (-1)^{k-1} s^{k-1} a^k \ . 
\label{ftest_taylor}
\eeq
As is evident from Eq. (\ref{ftest_taylor}), given that $s > 0$, the
coefficients of the $a^k$ terms in
the sum $\sum_{k=1}^\infty (-1)^{k-1} s^{k-1} a^k$ alternate in sign. This
property holds, independent of whether $r$ is zero or nonzero and, in the
latter case, independent of the sign of $r$.  The additional condition that $s
> r$ guarantees that the $O(a)$ term is opposite in sign from the constant
term, and hence that the full series is alternating in sign.  The resultant
alternating-sign property of the terms in the Taylor series
(\ref{ftest_taylor}) reproduces the alternating-sign property of the Taylor
series expansion of $\beta_{a,{\rm red.}}$ which holds for a large range of
values of $N$, namely $1 \le N \le 504$. Recall that all of the Pad\'e
approximants that we have calculated for $\beta_{a,n\ell,{\rm red.}}$ up to
$n=6$ loop order that have $p \ne 0$ and hence have zeros, also have the
property that they contain a pole closer to the origin than the zero of minimal
magnitude. This property is incorporated in the test function (\ref{ftest}),
since we take $s > r$.  Then $f(a) = 1-(s-r) a [1 - sa + (sa)^2 - (sa)^3 +
...]$.

As was discussed in the text, one of the purposes of our analysis with Pad\'e
approximants was to test for a stable zero and/or pole in $\beta_{a,{\rm
    red.}}$.  This Pad\'e method has the capability of doing this, as is
evident from the fact that when one calculates $[p,q]$ Pad\'e approximants to
the series (\ref{ftest_taylor}) with $p \ge 1$ and $q \ge 1$, they successfully
identify the exact function, $f(a)$, given in (\ref{ftest}).


\end{appendix}


\bigskip
\bigskip
\bigskip

\newpage

\begin{widetext}
\begin{table}
  \caption{ \footnotesize{Values of the $\bar b_\ell$ coefficients for
      $1 \le \ell \le 6$
      as functions of $N$ for $1 \le N \le 10$ and illustrative larger values
      of $N$. Notation $a$ \ e \ $n$ means $a \times 10^n$.}}
\begin{center}
\begin{tabular}{|c|c|c|c|c|c|c|} \hline\hline
$N$ & $\bar b_1$ & $\bar b_2$ & $\bar b_3$ & $\bar b_4$ & $\bar b_5$ &
$\bar b_6$
\\ \hline
1   & 0.2387   & $-0.03588$ & 0.01640 & $-0.01089$ & 0.09090 & $-0.008831$ \\
2   & 0.2653   & $-0.04222$ & 0.02013 & $-0.01406$ & 0.01227 & $-0.012443$ \\
3   & 0.2918   & $-0.04855$ & 0.02401 & $-0.01755$ & 0.01595 & $-0.016822$ \\
4   & 0.3183   & $-0.05488$ & 0.02805 & $-0.02137$ & 0.02016 & $-0.022035$ \\
5   & 0.3448   & $-0.06121$ & 0.03224 & $-0.02553$ & 0.02492 & $-0.028147$ \\
6   & 0.3714   & $-0.06755$ & 0.03658 & $-0.03001$ & 0.03024 & $-0.035229$ \\
7   & 0.3979   & $-0.07388$ & 0.04108 & $-0.03482$ & 0.03616 & $-0.043347$ \\
8   & 0.4244   & $-0.08021$ & 0.04573 & $-0.03996$ & 0.04269 & $-0.052571$ \\
9   & 0.4509   & $-0.08655$ & 0.05054 & $-0.04542$ & 0.04984 & $-0.062971$ \\
10  & 0.4775   & $-0.09288$ & 0.05550 & $-0.05121$ & 0.05765 & $-0.074616$ \\
30  & 1.00798  & $-0.21953$ & 0.18703 & $-0.23539$ & 0.38036 & $-0.682937$ \\
100 & 2.8648   & $-0.6628$  & 1.1324  & $-1.87505$ & 5.4152  & $-16.57724$ \\
200 & 5.5174   & $-1.2961$  & 3.7918  & $-6.7359$  & 26.0096 & $-1.28518$e2 \\
300 &  8.1700  & $-1.9293$  & 7.9910  & $-14.2812$ & 54.2973 & $-4.24105$e2 \\
400 & 10.8225  & $-2.5626$  & 13.7300 & $-24.2014$ & 63.0752 & $-0.932587$e3 \\
500 & 13.4751  & $-3.1958$  & 21.0087 & $-36.1873$ & 5.42998 & $-1.560139$e3 \\
600 & 16.1277  & $-3.8291$  & 29.8273 & $-49.9293$&$-1.85262$e2&$-2.02581$e3\\
700 & 18.7803  & $-4.4624$  & 40.1856 & $-65.1180$&$-5.95335$e2&$-1.79749$e3\\
800 & 21.4329  & $-5.0956$  & 52.0837 & $-81.4440$ &$-1.33083$e3&$-27.8255$ \\
900 & 24.0854  & $-5.7289$  & 65.5216 & $-98.5980$ &$-2.51752$e3&4.50979e3 \\
1.0e3&26.7380 & $-6.3621$  & 80.4992 & $-1.16270$e2 &$-4.30084$e3& 1.34853e4 \\
2.0e3&53.2639 & $-12.6947$ & 3.149645e2& $-2.53435$e2&$-1.01045$e5& 1.15991e6\\
3.0e3&79.7897 & $-19.0273$& 7.03408e2 & $-1.02078$e2& $-5.63816$e5&1.03446e7 \\
4.0e3&1.063155e2& $-25.3598$& 1.24583e3& 6.472275e2& $-1.86330$e6& 4.65918e7 \\
\hline\hline
\end{tabular}
\end{center}
\label{bbar_nloop_values}
\end{table}
%

\begin{table}
\caption{\footnotesize{Values of the UV zero $a_{_{UV,n\ell}}$ of the $n$-loop
beta function, $\beta_{\lambda,n\ell}$, for $n=2,...,6$, as a function of $N$,
with $b_n$ calculated in the $\overline{\rm MS}$
scheme for $3 \le n \le 6$. 
The notation ``u'' means that $\beta_{\lambda,n\ell}$ has only
unphysical (complex and/or negative) zeros for $a \ne 0$.}}
\begin{center}
\begin{tabular}{|c|c|c|c|c|c|c|} \hline\hline
$N$ & $a_{_{UV,2\ell}}$ & $a_{_{UV,3\ell}}$ & $a_{_{UV,4\ell}}$ &
$a_{_{UV,5\ell}}$ & $a_{_{UV,6\ell}}$
\\ \hline
1        &  0.5294   &  u   &  0.2333   &  u         & 0.1604  \\
2        &  0.5000   &  u   &  0.2217   &  u         & 0.1529  \\
3        &  0.4783   &  u   &  0.2123   &  u         & 0.1467  \\
4        &  0.4615   &  u   &  0.2044   &  u         & 0.1414  \\
5        &  0.4483   &  u   &  0.1978   &  u         & 0.1368  \\
6        &  0.4375   &  u   &  0.1920   &  u         & 0.1328  \\
7        &  0.4286   &  u   &  0.1869   &  u         & 0.1292  \\
8        &  0.42105  &  u   &  0.1823   &  u         & 0.1259  \\
9        &  0.4146   &  u   &  0.1783   &  u         & 0.1229  \\
10       &  0.4091   &  u   &  0.1746   &  u         & 0.1202  \\
30       &  0.3654   &  u   &  0.1362   &  u         & 0.09033 \\
100      &  0.3439   &  u   &  0.1012   &  u         & 0.05965 \\
300      &  0.3370   &  u   &  0.07944  &  u         & 0.03783 \\
500      &  0.3355   &  u   &  0.07341  &  0.08045   & 0.03074 \\
800      &  0.3347   &  u   &  0.07137  &  0.02871   & 0.02866 \\
890      &  0.3346   &  u   &  0.07164  &  0.02559   & 0.03829 \\
900      &  0.3346   &  u   &  0.07170  &  0.02530   & u       \\
1000     &  0.3344   &  u   &  0.07241  &  0.02276   & u       \\
2000     &  0.3339   &  u   &  0.1054   &  0.01231   & u       \\
3000     &  0.3337   &  u   &  0.5475   &  0.008850  & u       \\
4000     &  0.3336   &  u   &  u        &  0.007042  & u       \\
$10^4$   &  0.3334   &  u   &  u        &  0.003460  & u       \\
\hline\hline
\end{tabular}
\end{center}
\label{auv_nloop_values}
\end{table}


\begin{table}
  \caption{\footnotesize{Values of real zeros and poles in the $[p,q]$ Pad\'e
      approximants to the $n$-loop reduced beta function, 
      $\beta_{a,n\ell,{\rm red.}}$ for $2 \le n \le 6$ and $N=1$, with $b_n$, 
      $3 \le n \le 6$, calculated in the $\overline{\rm MS}$ scheme. Note 
      that the $[n-1,0]$ Pad\'e approximant is the function 
      $\beta_{a,n\ell,{\rm red.}}$ itself, whose zeros are given in Table 
      \ref{auv_nloop_values}. The symbol ``na'' means ``not applicable'', and 
      the symbols ``ccp'' and $k(ccp)$ mean a complex-conjugate pair of 
values and $k$ complex-conjugate pairs of values, respectively.}}
\begin{center}
\begin{tabular}{|c|c|c|c|} \hline\hline
$n$ & $[p,q]$ & zeros & poles 
\\ \hline
2 & [0,1] & na & $-0.5294$  \\
\hline
3 & [1,1] & $-0.2594$  & $-0.1741$ \\
3 & [0,2] & na & $-0.2629$, \ $0.5223$ \\
\hline
4 & [2,1] & $-0.1400$, \ $1.4543$ & $-0.1198$ \\
4 & [1,2] & $-0.1294$  & $-0.1138$, $-1.2005$  \\
4 & [0,3] & na         & $-0.1893$, \ ccp  \\
\hline
5 & [3,1] & $-0.1024$, \ ccp & $-0.09535$ \\
5 & [2,2] & $-0.08736$, \ $-0.5298$ & $-0.08401$, \ $-0.3013$ \\
5 & [1,3] & $-0.09489$ & $-0.08986$, \ $-0.4644$, \ 1.1714 \\
5 & [0,4] & na         & $-0.1538$, \ 0.2334, \ ccp \\
\hline
6 & [4,1] & $-0.085055$, \ 0.4675, \ ccp & $-0.08191$  \\
6 & [3,2] & $-0.07366$, \ $-0.2637$, \ 2.8463  & $-0.07233$, \ $-0.1968$  \\
6 & [2,3] & $-0.07279$, \ $-0.2430$ & $-0.07156$, \ $-0.1878$, \ $-2.2495$ \\
6 & [1,4] & $-0.08007$ & $-0.07784$, \ $-0.3012$, \ ccp  \\
6 & [0,5] & na & $-0.1327$, \ 2(ccp) \\
\hline\hline
\end{tabular}
\end{center}
\label{pade_zp_Neq1}
\end{table}


\begin{table}
  \caption{\footnotesize{Values of real zeros and poles in the $[p,q]$ Pad\'e
   approximants to the $n$-loop reduced beta function, 
$\beta_{a,n\ell,{\rm red.}}$ for $2 \le n \le 6$ and $N=10$, with 
$b_n$, $3 \le n \le 6$, calculated in the $\overline{\rm MS}$ scheme. 
Note that the $[n-1,0]$ Pad\'e approximant is the function 
$\beta_{a,n\ell,{\rm red.}}$ itself, whose zeros are given in Table 
\ref{auv_nloop_values}. The notation is the same as in Table 
\ref{pade_zp_Neq1}.}}
\begin{center}
\begin{tabular}{|c|c|c|c|} \hline\hline
$n$ & $[p,q]$ & zeros & poles 
\\ \hline
2 & [0,1] & na & $-4091$  \\
\hline
3 & [1,1] & $-0.1974$  & $-0.1332$ \\
3 & [0,2] & na & $-0.2021$, \ $0.3996$  \\
\hline
4 & [2,1] & $-0.09864$, \ 1.01465 &  $-0.08623$  \\
4 & [1,2] & $-0.08989$   & $-0.0808$, \ $-0.8352$  \\
4 & [0,3] & na    & $-0.1426$, \ ccp  \\
\hline
5 & [3,1] & $-0.07576$, \ ccp & $-0.07069$ \\
5 & [2,2] & $-0.06870$, \ $-0.5462$  & $-0.06547$, \ $-0.2812$ \\
5 & [1,3] & $-0.071475$  & $-0.06757$, \ $-0.3892$, \ 1.0716  \\
5 & [0,4] & na       & $-0.1154$, \ 0.1743, \ ccp \\
\hline
6 & [4,1] & $-0.06388$, \ 0.3526, \ ccp & $-0.06149$  \\
6 & [3,2] & $-0.05486$, \ $-0.1701$, \ 1.5187 & $-0.05391$, \ $-0.1362$  \\
6 & [2,3] & $-0.05259$, \ $-0.1412$ & $-0.05188$, \ $-0.1194$, \ $-1.1224$ \\
6 & [1,4] & $-0.06071$ & $-0.05892$, \ $-0.2352$, \ ccp \\
6 & [0,5] & na          & $-0.0099505$, \ 2(ccp) \\
\hline\hline
\end{tabular}
\end{center}
\label{pade_zp_Neq10}
\end{table}
\end{widetext}



\begin{thebibliography}{99}
\bibitem{nintegral}
%
Here and below, when expressions are given for $N$ that evaluate
to non-integral real values, it is understood that they are formal and are
interpreted via an analytic continuation of $N$ from physical nonnegative
integer values to real numbers.
\end{thebibliography}
\end{document}